\documentclass[journal]{IEEEtran}
\makeatletter
\def\ps@IEEEtitlepagestyle{%
  \def\@oddfoot{\mycopyrightnotice}%
  \def\@oddhead{\hbox{}\@IEEEheaderstyle\leftmark\hfil\thepage}\relax
  \def\@evenhead{\@IEEEheaderstyle\thepage\hfil\leftmark\hbox{}}\relax
  \def\@evenfoot{}%
}
\def\mycopyrightnotice{%
  \begin{minipage}{\textwidth}
  \centering \scriptsize
  Copyright~\copyright~2022 IEEE. Personal use of this material is permitted. Permission from IEEE must be obtained for all other uses, in any current or future media, including\\reprinting/republishing this material for advertising or promotional purposes, creating new collective works, for resale or redistribution to servers or lists, or reuse of any copyrighted component of this work in other works.
  \end{minipage}
}
\makeatother
\usepackage{caption}
\usepackage{subcaption}

\ifCLASSINFOpdf
\else
   \usepackage[dvips]{graphicx}
\fi
\usepackage{url}
\usepackage{array}
\usepackage{hyperref}
\usepackage{graphicx}
\usepackage{booktabs}
\usepackage{balance}
\usepackage{amsmath}
\usepackage{amsfonts}
\usepackage{multirow}
\usepackage{cite}
\begin{document}

\title{SNAC: Speaker-normalized affine coupling layer in flow-based architecture for zero-shot multi-speaker text-to-speech}
\author{Byoung Jin Choi,~\IEEEmembership{Student Member, IEEE},
        Myeonghun Jeong,~\IEEEmembership{Student Member, IEEE},
        Joun Yeop Lee, \\
        and Nam Soo Kim,~\IEEEmembership{Senior Member, IEEE}
\thanks{This work was supported by Samsung Research, Samsung Electronics Co.,Ltd.

Byoung Jin Choi, Myeonghun Jeong, and Nam Soo Kim are with the Department of Electrical and Computer Engineering and with the Institute of New Media and Communications, Seoul National University, Seoul 08826, South Korea (e-mail: bjchoi@hi.snu.ac.kr; mhjeong@hi.snu.ac.kr; nkim@snu.ac.kr)

Joun Yeop Lee is with Samsung Research, Samsung Electronics Co., Ltd, Seoul, South Korea (e-mail: jounyeop.lee@samsung.com)}}

\markboth{}
{Shell \MakeLowercase{\textit{et al.}}: Bare Demo of IEEEtran.cls for IEEE Journals}
\maketitle

\begin{abstract}
Zero-shot multi-speaker text-to-speech (ZSM-TTS) models aim to generate a speech sample with the voice characteristic of an unseen speaker. The main challenge of ZSM-TTS is to increase the overall speaker similarity for unseen speakers. One of the most successful speaker conditioning methods for flow-based multi-speaker text-to-speech (TTS) models is to utilize the functions which predict the scale and bias parameters of the affine coupling layers according to the given speaker embedding vector. In this letter, we improve on the previous speaker conditioning method by introducing a speaker-normalized affine coupling (SNAC) layer which allows for unseen speaker speech synthesis in a zero-shot manner leveraging a normalization-based conditioning technique. The newly designed coupling layer explicitly normalizes the input by the parameters predicted from a speaker embedding vector while training, enabling an inverse process of denormalizing for a new speaker embedding at inference. The proposed conditioning scheme yields the state-of-the-art performance in terms of the speech quality and speaker similarity in a ZSM-TTS setting.

\end{abstract}


\IEEEpeerreviewmaketitle

\section{Introduction}

\IEEEPARstart{A}{s} the sample quality of the recently proposed neural text-to-speech (TTS) models, \cite{wang2017tacotron, shen2018natural, tachibana2018efficiently, arik2017deep, ping2017deep, li2019neural, kim2021conditional, ren2019fastspeech, ren2020fastspeech, kim2020glow, donahue2020end}, approaches to the natural speech, the research interest has extended to high fidelity multi-speaker TTS systems, which enables the speech generation of multiple speakers via a single trained model. However, training a multi-speaker TTS system requires a large dataset of \([text, audio, speaker]\) tuples where the labeling can be costly. Furthermore, such systems are limited to generate the voice of speakers seen during the training period, whereas instant adaptation to a new speaker's voice may be required in real life applications. To this end, \emph{personalized TTS} is gaining huge attention from researchers.

Personalized TTS aims at generating new speakers' speech with limited resources. One possible approach is speaker adaptation. The idea of adapting a pre-trained TTS model to a new speaker with more than one \([text, audio]\) pair dates back to the hidden Markov model (HMM)-based TTS \cite{yoshimura1999simultaneous,tokuda1995speech,masuko1996speech,masuko1997voice}. \cite{tamura1998speaker} and \cite{tamura2001adaptation} extend the maximum likelihood linear regression (MLLR) algorithm for speaker adaptation. For more robust speaker adaptation, structured maximum \emph{a posteriori} linear regression (SMAPLR) \cite{siohan2002structural} is developed by combining the maximum \emph{a posteriori} (MAP) criterion to MLLR. The adaptation process is based on the affine transformation of mean and variance of HMM parameters for the target speaker and such transformation matrices are derived by maximum likelihood and MAP criterion respectively. With the recent development of non-autoregressive neural TTS systems, \cite{arik2018neural} and \cite{chen2021adaspeech} focus on effectively finetuning the parameters of pre-trained neural TTS model to adapt to the speaker's characteristics.

Another approach deals with an extreme situation where only an \([audio]\) from a target speaker is available. The model is required to correctly reflect the unseen target speaker's characteristics without further finetuning the model. This task is known as zero-shot multi-speaker TTS (ZSM-TTS). Some of the previous works, \cite{jia2018transfer,cooper2020zero,casanova2021sc,casanova2022yourtts}, propose using an external speaker encoder trained for speaker verification. \cite{min2021meta,lee2021multi,choi2022adversarial} utilized adversarial training to enhance unseen speaker generalization. On the other hand, normalization-based conditioning techniques used in style transfer, \cite{ulyanov2016instance, huang2017arbitrary}, were introduced to condition speaker embeddings for FastSpeech-based models in \cite{min2021meta} and \cite{kumar2021normalization}. These conditioning methods first remove the instance-specific information from the input to preserve content via speaker-normalization. The normalized input is then scaled and shifted by the affine parameters predicted from the target speaker embedding vector.

However, recently proposed flow-based TTS models are rather under-explored in ZSM-TTS applications. Leveraging the aforementioned normalization-based speaker conditioning techniques in flow-based models is especially challenging because the flow requires the inverse operation of such normalization unlike feed-forward models.

In this letter, we propose a speaker-normalized affine coupling (SNAC) layer for flow-based TTS models in the ZSM-TTS scenario. The proposed method explicitly normalizes the input by the speaker-dependent parameters in order to preserve speaker-independent information at training, while target speaker's information is inversely injected at inference through denormalization. We compare the proposed conditioning method to the existing method in several different experimental settings using VITS \cite{kim2021conditional} as our base model and demonstrate that the proposed method outperforms the conventional technique in both subjective and objective measures. The audio samples are available on the demo page\footnote{\url{https://byoungjinchoi.github.io/snac/}}.

\section{Affine coupling-based generative flow}
Normalizing flow models, \cite{dinh2014nice, dinh2016density, kingma2018glow}, learn an invertible mapping between a prior distribution, \(p_\theta(z)\), and a more complex data distribution \(p_\theta(x)\) using a sequence of bijective functions. The log-likelihood computation is tractable via the rule of change-of-variables. Let \(f_\theta : \mathbb{R}^{D}\rightarrow{}\mathbb{R}^{D}\) be a bijective function which maps the observed data \(x\) to the latent variable \(z\) from a simple prior distribution \(p_\theta(z)\) where \(x, z \in \mathbb{R}^{D}\). Then the log-likelihood is obtained by
\begin{equation}\label{equation1}
    \log p_\theta(x)=\log p_\theta(z) + \log \left|\text{det}\dfrac{{\partial}f_\theta(x)}{{\partial}x}\right|_{\textstyle .}
\end{equation}
Computing the log determinant of the Jacobian matrix in (\ref{equation1}) is computationally expensive in general. In addition, \(f_\theta\) is strictly restricted to be a bijective function in which only certain types of transformations can be easily inversed. An affine coupling layer, first introduced in \cite{dinh2014nice}, allows for an efficient computation of the log determinant with invertible transformations by generating an output \(y \in \mathbb{R}^{D}\) given an input \(x \in \mathbb{R}^{D}\) and \(d < D\) via
\begin{gather}\label{equation2}
\begin{gathered}
    y_{1:d} = x_{1:d} \\
    y_{d+1:D} = x_{d+1:D}\,{\odot}\,\exp({s_\theta(x_{1:d})}) + b_\theta(x_{1:d})
\end{gathered}
\end{gather}
 where \(s_\theta\) and \(b_\theta\) are parameterized scale and bias functions mapping \(\mathbb{R}^{d}\rightarrow{}\mathbb{R}^{D-d}\), and \(\odot\) is an element-wise product. With this coupling architecture, the Jacobian becomes a lower triangular matrix as given by
\begin{gather}\label{equation3}
    \frac{\partial y}{\partial x} = 
    \begin{bmatrix}
        \mathbb{I}_d & 0 \\
        \frac{\partial y_{d+1:D}}{\partial x_{1:d}} & \text{diag}(\exp({s_\theta(x_{1:d})}))
    \end{bmatrix}
\end{gather}
where \(\mathbb{I}_d\) represents a \(d \times d\) identity matrix.
Computing the determinant of Jacobian matrix of the affine coupling does not depend on the Jacobians of \(s_\theta\) and \(b_\theta\). Therefore, they can be any type of complex functions modeled by highly expressive neural networks, such as non-causal Wavenet \cite{oord2016wavenet}. The inverse transformation of the coupling layer can be easily derived as
\begin{gather}\label{equation4}
\begin{gathered}
    x_{1:d} = y_{1:d} \\
    x_{d+1:D} = \dfrac{y_{d+1:D} - b_\theta(y_{1:d})}{\exp({s_\theta(y_{1:d})})}
\end{gathered}_{\textstyle ,}
\end{gather}
hence sampling is also efficient. Each coupling layer is then followed by a layer which permutes the ordering of the channels along the feature dimension.

\section{Speaker-normalized affine coupling layer for ZSM-TTS}
A conditional generative flow \cite{atanov2019semi, serra2019blow} models a conditional probability distribution \(p_\theta(x|g)\) where \(g\) represents a conditioning term. Conventionally, a conditional flow extends the forward and inverse transformations of an affine coupling layer given in (\ref{equation2}) and (\ref{equation4}) by modifying \(s_\theta\) and \(b_\theta\) such that they take \(g\) as an additional input.

For ZSM-TTS, the condition \(g\) usually represents a specific speaker embedding vector. Our strategy for ZSM-TTS is to convert the speaker-dependent data distribution to a latent prior distribution which becomes speaker-independent. Then, when synthesizing speech, the speaker-independent latent prior distribution is mapped back to a speaker-specific data distribution depending on the given speaker embedding. In order to achieve this, we design each affine coupling layer to remove the information related to \(g\) at the forward transformation. Contrarily, \(g\) is injected to the input embedding sequence at the inverse transformation. To obtain such bijective transformation with explicit \(g\) conditioning, we propose a speaker-normalized affine coupling (SNAC) layer, which normalizes and denormalizes the input embedding sequence by the mean and standard deviation parameters predicted by \(g\). Speaker normalization (\(SN\)) and speaker denormalization (\(SDN\)) in SNAC are performed as follows:
\begin{gather}\label{equation5}
\begin{gathered}
    SN(x;g) = \dfrac{x - m_{\theta}(g)}{\exp{(v_\theta(g))}} \\
    SDN(x;g) = x\,{\odot}\,\exp{(v_{\theta}(g))} + m_{\theta}(g)
\end{gathered}
\end{gather}
where \(m_{\theta}\) and \(v_{\theta}\) are simple linear projections to obtain the mean and standard deviation parameters from \(g\). \(SN\) and \(SDN\) are applied across the temporal axis, thus normalizing and denormalizing each frame of the input \(x\) with the same mean and standard deviation parameters.

The forward transformation of the SNAC layer is now given by 
\begin{gather}\label{equation6}
\begin{gathered}
    y_{1:d} = x_{1:d} \\
    y_{d+1:D} = SN(x_{d+1:D};g)\,{\odot}\,\exp({s_\theta(SN(x_{1:d};g))})\\
    + b_\theta(SN(x_{1:d};g))
\end{gathered}_{\textstyle .}
\end{gather}

The inverse transformation can be derived straightforwardly as follows:
\begin{gather}\label{equation7}
\begin{gathered}
    x_{1:d} = y_{1:d} \\
    x_{d+1:D} = SDN\left(\dfrac{y_{d+1:D} - b_\theta(SN(y_{1:d};g))}{\exp({s_\theta(SN(y_{1:d};g))})};g\right) \\
\end{gathered}_{\textstyle .}
\end{gather}

At each SNAC layer, \(SN\) is applied to the input of \(s_{\theta}\) and \(b_{\theta}\) such that the affine parameters contain the information unrelated to the speaker. Since \(x_{d+1:D}\) is also speaker-normalized in the forward transformation, this results in the extensive removal of speaker information at training. When inferring \(x_{d+1:D}\) through the inverse transformation, \(SDN\) is enforced after the affine transformation is applied to \(y_{d+1:D}\) to appropriately inject information related to the target speaker.

\begin{figure}
\centering
\begin{subfigure}{0.47\columnwidth}
\includegraphics[width=\linewidth]{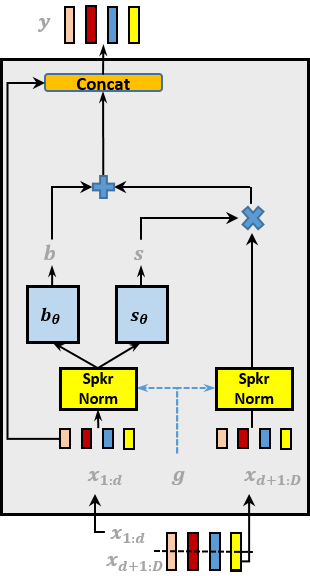}
\caption{Forward transformation}
\label{fig:subim_train}
\end{subfigure}
\begin{subfigure}{0.47\columnwidth}
\includegraphics[width=\linewidth]{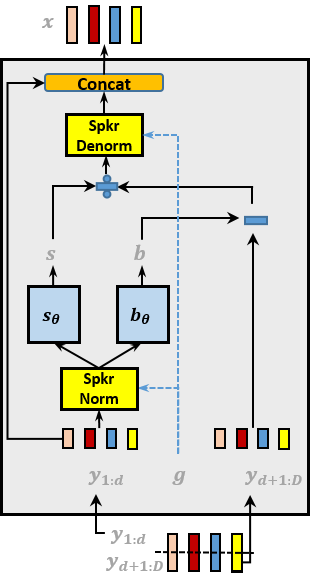}
\caption{Inverse transformation}
\label{fig:subim_infer}
\end{subfigure}
\caption{(a) and (b) each show the forward and inverse transformations of SNAC layer.}
\label{fig:image1}
\end{figure}

The log-determinant of the conditional flow with SNAC layers can be obtained by
\begin{gather}\label{equation8}
    \log \left|\text{det}\dfrac{{\partial}f_{\theta}(x)}{{\partial}x}\right| = \log \sum_j\dfrac{\exp({s_\theta(SN(x_{1:d};g))_j})}{\exp{(v_\theta(g)_j)}}_{\textstyle .}
\end{gather}
The complete architecture of the SNAC layer is presented in Fig. \ref{fig:image1}.

\section{Experiments}
\subsection{Model}
We performed experiments on the proposed method by replacing the affine coupling layer of the flow module in VITS with the SNAC layer.
\subsubsection{VITS overview}
VITS leverages the variational autoencoder (VAE) \cite{kingma2013auto} formulation and the adversarial training strategy to successfully combine the joint training of acoustic feature generation, vocoding, and duration prediction in an end-to-end manner. The objective is to maximize the variational lower bound of a conditional log-likelihood
\begin{equation}\label{equation9}
    \log p_{\theta}(o|l) \geq E_{q_{\phi}(z|o)}\left[\log p_{\theta}(o|z)-\log \dfrac{q_{\phi}(z|o)}{p_{\theta}(z|l)}\right]
\end{equation}
where \(p_{\theta}(o|z)\), \(q_{\phi}(z|o)\) and \(p_{\theta}(z|l)\) respectively denote the likelihood, the approximate posterior and the conditional prior distributions. A target speech and the corresponding phoneme sequence are denoted as \(o\) and \(l=[l_{text}, A]\), and \(z\) is a frame-level latent sequence representing the intermediate acoustic features. The alignment \(A\) is estimated using the monotonic alignment search (MAS) algorithm proposed in \cite{kim2020glow}. The generator part of VITS architecture consists of a posterior encoder, prior encoder, decoder, and duration predictor and is trained with a discriminator in an adversarial manner. The prior encoder is composed of two parts: a text encoder and a flow module. The flow module plays an essential role in transforming a simple text-conditional distribution to a more complex distribution.

\subsubsection{Multi-speaker VITS}
In a multi-speaker setting, the likelihood \(p_\theta(o|l)\) is substituted with \(p_\theta(o|l,g)\) where \(g\) represents a speaker embedding. For training, the original work uses a speaker label as an additional input which is transformed to a fixed-dimensional vector \(g\) via a learnable embedding table. \(g\) is then conditioned to every module of the generator.

\subsection{Datasets}
All tested models were trained on VCTK \cite{vctk} dataset. VCTK is a multi-speaker audio dataset which contains approximately 44 hours of speech recorded by 109 speakers. We selected 11 speakers as an in-domain test set following \cite{casanova2022yourtts}. To evaluate the performance on the out-of-domain dataset, we randomly selected 20 speakers from LibriTTS \cite{zen2019libritts} test-clean dataset, which consists of 8.56 hours of audio from 39 speakers. Each utterance was downsampled to 22050 Hz for training.

\subsection{Implementation details}
Our proposed method modifies the official implementation of VITS\footnote{\url{https://github.com/jaywalnut310/vits}}. For the partitioning scheme of affine coupling layer at flow module, we chose channel-wise masking pattern \cite{dinh2016density}. To ensure that all input entries are processed, we reverse the ordering of the feature dimension at each layer of flow module. We employ a reference encoder to extract the speaker embedding vector. The reference encoder is composed of a stack of 2-D convolutional layers and a gated recurrent unit (GRU) \cite{chung2014empirical}, following global style token (GST) \cite{wang2018style}. The reference encoder takes a sequence of linear spectrograms of the reference audio as an input and outputs a 256-dimensional embedding vector. Two separate linear projection layers, \(m_{\theta}\) and \(v_\theta\), are employed to predict the mean and standard deviation parameters of the speaker in (\ref{equation5}) from the reference embedding vector. 

\begin{table*}[htb!]
    \renewcommand{\arraystretch}{1.15}%
    \centering
    \begin{tabular}{c c c c | c c c}
        \toprule
        \multirow{2}{*}{\textbf{Model}}&\multicolumn{3}{c}{\textbf{VCTK}}&\multicolumn{3}{c}{\textbf{LibriTTS}}\\
        \cline{2-7} & \textbf{MOS}($\uparrow$) & \textbf{SMOS}($\uparrow$) & \textbf{SECS}($\uparrow$) & \textbf{MOS}($\uparrow$) & \textbf{SMOS}($\uparrow$) & \textbf{SECS}($\uparrow$)\\
        \specialrule{0.1pt}{1pt}{1pt}
        Ground Truth & 4.76$\pm$0.02 & 4.19$\pm$0.04 & 0.748 & 4.80$\pm$0.02 & 4.51$\pm$0.03 & 0.646\\
        \specialrule{0.1pt}{1pt}{1pt}
        Meta-StyleSpeech & 2.06$\pm$0.04 & 2.62$\pm$0.05 & 0.212 & 2.00$\pm$0.03 & 2.50$\pm$0.04 & 0.131\\
        YourTTS & 4.42$\pm$0.03 & 3.86$\pm$0.04 & \textbf{0.447} & 4.23$\pm$0.03 & 3.35$\pm$0.04 & \textbf{0.317}\\
        Baseline+REF+ALL & 4.22$\pm$0.04 & 4.11$\pm$0.04 & 0.350 & 4.30$\pm$0.03 & 3.67$\pm$0.04 & 0.143\\
        Baseline+REF+FLOW & 4.08$\pm$0.04 & 4.01$\pm$0.04 & 0.339 & 3.98$\pm$0.04 & 3.64$\pm$0.04 & 0.135\\
        Baseline+PRE-TRAINED+FLOW & 4.38$\pm$0.03 & 3.52$\pm$0.04 & 0.321 & 4.17$\pm$0.03 & 2.91$\pm$0.05 & 0.135\\
        \specialrule{0.1pt}{1pt}{1pt}
        Proposed+REF+ALL & 4.30$\pm$0.03 & 4.07$\pm$0.04 & 0.320 & 4.11$\pm$0.03 & 3.56$\pm$0.04 & 0.145\\
        \textbf{Proposed+REF+FLOW} & \textbf{4.48$\pm$0.03} & \textbf{4.19$\pm$0.04} & 0.352 & \textbf{4.41$\pm$0.03} & \textbf{3.70$\pm$0.04} & 0.151\\
        Proposed+PRE-TRAINED+FLOW & 4.46$\pm$0.03 & 3.61$\pm$0.04 & 0.270 & 4.40$\pm$0.03 & 3.18$\pm$0.04 & 0.116\\
        \bottomrule
    \end{tabular}
    \caption{MOS, SMOS, and SECS on unseen speakers of VCTK and LibriTTS}
    \label{table:1}
\end{table*}

\subsection{Experimental setup}
We evaluated our method in several different settings. We built our first baseline, \textbf{Baseline+REF+ALL}, by attaching a reference encoder to the vanilla multi-speaker VITS model, which applied the speaker conditioning at every module of the generator. The second baseline, \textbf{Baseline+REF+FLOW}, conditioned the speaker embedding only to the duration predictor and the flow module to focus on the effect of the proposed method. The last baseline, \textbf{Baseline+PRE-TRAINED+FLOW}, substituted the reference encoder with a pre-trained speaker encoder which was trained from a speaker verification for a different speaker embedding scenario. We used a H/ASP model \cite{heo2020clova} which can be obtained from an open-source project\footnote{\url{https://github.com/clovaai/voxceleb_trainer}}. In this baseline, the pre-trained speaker encoder weights were fixed to consistently draw speaker embedding vectors from a learned speaker embedding space. For the above three baselines, the speaker embedding vector was used as a conditional input to produce \(s_{\theta}\) and \(b_{\theta}\) at affine coupling layers of the flow module.

To demonstrate the effect of the proposed method for each setting, we replaced the conventional affine coupling layers with the SNAC layers for the above three baselines. We name the three proposed models corresponding to each baseline as follows: \textbf{Proposed+REF+ALL}, \textbf{Proposed+REF+FLOW}, \textbf{Proposed+PRE-TRAINED+FLOW}.

Furthermore, we compared our models with two other baseline models: \textbf{Meta-StyleSpeech} \cite{min2021meta} and \textbf{YourTTS} \cite{casanova2022yourtts}. \textbf{Meta-StyleSpeech} is trained with a meta-learning scheme with a modified structure of FastSpeech2 \cite{ren2020fastspeech}. \textbf{YourTTS} is built on VITS architecture with an external speaker encoder and an additional speaker consistency loss. We used an open-source implementation\footnote{\url{https://github.com/keonlee9420/StyleSpeech}} for \textbf{Meta-StyleSpeech} and followed the paper to implement \textbf{YourTTS} on the official VITS code.

\subsection{Evaluation method}
We first conducted subjective tests to measure the overall speech quality using mean opinion score (MOS). To assess the effectiveness of the proposed speaker conditioning method, we also measured similarity mean opinion score (SMOS). SMOS is employed to evaluate how similar the synthesized samples are to the reference speech samples in terms of speaker characteristic. Both MOS and SMOS are 5-scale scores ranging from 1 to 5 and reported with the 95\% confidence interval. For in-domain evaluations, we drew 3 pairs of text and reference audio randomly from each of the 11 unseen speakers of VCTK test dataset. To evaluate on the out-of-domain case, 2 pairs of text and reference audio from 20 randomly selected speakers were drawn from the LibriTTS test-clean dataset. 15 judges participated in the subjective tests with both VCTK and LibriTTS unseen speakers.

In addition, we also evaluated the objective score for speaker similarity between the generated samples and the ground truth samples using speaker embedding cosine similarity (SECS). The SECS ranges from -1 to 1, where 1 indicates both samples are from the same speaker. We computed SECS using a pre-trained speaker encoder model provided by SpeechBrain toolkit\footnote{\url{https://github.com/speechbrain/speechbrain}} \cite{ravanelli2021speechbrain}. The results of MOS, SMOS, and SECS are presented in Table \ref{table:1}. 
\setlength{\tabcolsep}{1pt}

\subsection{Results}
The MOS and SMOS results shown in Table \ref{table:1} indicate that \textbf{Proposed+REF+FLOW} consistently shows superior performance over the baseline models in terms of sample quality and speaker similarity.

In \textbf{Proposed+REF+ALL}, the SNAC-based flow module enforces the explicit removal of speaker information at the forward transformation while the speaker information is conversely injected to the generator modules. Training this model results in neutralizing the effect of the SNAC layer, and this accounts for the MOS and SMOS drop from \textbf{Baseline+REF+ALL} to \textbf{Proposed+REF+ALL} in LibriTTS dataset. On the contrary, the best performance for subjective tests is consistently achieved by \textbf{Proposed+REF+FLOW} in both VCTK and LibriTTS datasets. Nonetheless, \textbf{YourTTS} shows the highest SECS scores among all models since \textbf{YourTTS} is trained to minimize the speaker embedding cosine similarity. 

From the synthesized samples, we have noticed that the models using pre-trained speaker encoders occasionally produce the voice of different speakers. This phenomenon is reflected in the lower SMOS of \textbf{Baseline+PRE-TRAINED+FLOW} and \textbf{Proposed+PRE-TRAINED+FLOW}. This shows that the joint training of a reference encoder may be more suitable for ZSM-TTS task than using a pre-trained speaker encoder in terms of speaker stability. However, this does not affect the MOS as much since the generated samples maintain the consistent quality.

Although \textbf{Proposed+REF+ALL} outperforms the baseline models in both in-domain and out-of-domain datasets, the overall performance drop between the two settings still exists in terms of speaker similarity. Since LibriTTS dataset inherently includes various types of channel conditions which may interfere with the accurate inference of speaker embedding while VCTK contains only clean speech data, we conjecture such domain mismatch accounts for the performance drop.

\section{Conclusion}
We have proposed a novel speaker conditioning method for flow-based multi-speaker TTS. The experimental results show that the proposed method outperforms the conventional conditioning technique in a ZSM-TTS setting and achieves the best performance in subjective tests compared to the other baseline models. For a future work, we intend to incorporate locally-varying features related to prosody and accents.

\section*{Acknowledgment}
This work was supported by Samsung Research, Samsung Electronics Co.,Ltd.

\bibliographystyle{ieeetr}
\bibliography{bibliography.bib}
\end{document}